\newif\iflanl
\openin 1 lanlmac
\ifeof 1 \lanlfalse \else \lanltrue \fi
\closein 1
\iflanl
    \input lanlmac
\else
    \message{[lanlmac not found - use harvmac instead}
    \input harvmac
\fi
\newif\ifhypertex
\ifx\hyperdef\UnDeFiNeD
    \hypertexfalse
    \message{[HYPERTEX MODE OFF}
    \def\hyperref#1#2#3#4{#4}
    \def\hyperdef#1#2#3#4{#4}
    \def\hypernoname{}
    \def\e@tf@ur#1{}
    \def\hth/#1#2#3#4#5#6#7{{\tt hep-th/#1#2#3#4#5#6#7}}
    \def\CERN{\address{CERN, Geneva, Switzerland}}
\else
    \hypertextrue
    \message{[HYPERTEX MODE ON}
  \def\hth/#1#2#3#4#5#6#7{
  {\tt hep-th/#1#2#3#4#5#6#7}}
\def\CERN{\address{

Theory Division, CERN, Geneva, Switzerland}}
\fi
\newif\ifdraft

\noblackbox
\catcode`\@=11
\newif\iffrontpage
\ifx\answ\bigans
\def\titleft{\titla}
\magnification=1200\baselineskip=14pt plus 2pt minus 1pt
%
\advance\hoffset by-0.075truein
\advance\voffset by1.truecm
\hsize=6.15truein\vsize=600.truept\hsbody=\hsize\hstitle=\hsize
\else\let\lr=L
\def\titleft{\titla}
\magnification=1000\baselineskip=14pt plus 2pt minus 1pt
%
\hoffset=-0.75truein\voffset=-.0truein
\vsize=6.5truein
\hstitle=8.truein\hsbody=4.75truein
\fullhsize=10truein\hsize=\hsbody
\fi
\parskip=4pt plus 15pt minus 1pt
%
\newif\iffigureexists
\newif\ifepsfloaded
\def\epsfcheck{
\ifdraft
\input epsf\epsfloadedtrue
\else
  \openin 1 epsf
  \ifeof 1 \epsfloadedfalse \else \epsfloadedtrue \fi
  \closein 1
  \ifepsfloaded
    \input epsf
  \else
\immediate\write20{NO EPSF FILE --- FIGURES WILL BE IGNORED}
  \fi
\fi
\def\epsfcheck{}}
\def\checkex#1{
\ifdraft
\figureexistsfalse\immediate%
\write20{Draftmode: figure #1 not included}
\else\relax
    \ifepsfloaded \openin 1 #1
        \ifeof 1
           \figureexistsfalse
  \immediate\write20{FIGURE FILE #1 NOT FOUND}
        \else \figureexiststrue
        \fi \closein 1
    \else \figureexistsfalse
    \fi
\fi}
\def\missbox#1#2{$\vcenter{\hrule
\hbox{\vrule height#1\kern1.truein
\raise.5truein\hbox{#2} \kern1.truein \vrule} \hrule}$}
\def\lfig#1{
\let\labelflag=#1%
\def\numb@rone{#1}%
\ifx\labelflag\UnDeFiNeD%
{\xdef#1{\the\figno}%
\writedef{#1\leftbracket{\the\figno}}%
\global\advance\figno by1%
}\fi{\hyperref{}{figure}{{\numb@rone}}{Fig.{\numb@rone}}}}
\def\figinsert#1#2#3#4{
\epsfcheck\checkex{#4}%
\def\figsize{#3}%
\let\flag=#1\ifx\flag\UnDeFiNeD
{\xdef#1{\the\figno}%
\writedef{#1\leftbracket{\the\figno}}%
\global\advance\figno by1%
}\fi
\goodbreak\midinsert%
\iffigureexists
\centerline{\epsfysize\figsize\epsfbox{#4}}%
\else%
\vskip.05truein
  \ifepsfloaded
  \ifdraft
  \centerline{\missbox\figsize{Draftmode: #4 not included}}%
  \else
  \centerline{\missbox\figsize{#4 not found}}
  \fi
  \else
  \centerline{\missbox\figsize{epsf.tex not found}}
  \fi
\vskip.05truein
\fi%
{\smallskip%
\leftskip 4pc \rightskip 4pc%
\noindent\ninepoint\sl \baselineskip=11pt%
{\bf{\hyperdef\hypernoname{figure}{{#1}}{Fig.{#1}}}:~}#2%
\smallskip}\bigskip\endinsert%
}

\def\boxit#1{\vbox{\hrule\hbox{\vrule\kern8pt
\vbox{\hbox{\kern8pt}\hbox{\vbox{#1}}\hbox{\kern8pt}}
\kern8pt\vrule}\hrule}}
\def\mathboxit#1{\vbox{\hrule\hbox{\vrule\kern8pt\vbox{\kern8pt
\hbox{$\displaystyle #1$}\kern8pt}\kern8pt\vrule}\hrule}}

%
\font\bigit=cmti10 scaled \magstep1

\font\titla=cmr10 scaled\magstep3
\font\tenmss=cmss10
\font\absmss=cmss10 scaled\magstep1

\newfam\mssfam
\font\footrm=cmr8  \font\footrms=cmr5
\font\footrmss=cmr5   \font\footi=cmmi8
\font\footis=cmmi5   \font\footiss=cmmi5
\font\footsy=cmsy8   \font\footsys=cmsy5
\font\footsyss=cmsy5   \font\footbf=cmbx8
\font\footmss=cmss8
\def\footfont{\def\rm{\fam0\footrm}
\textfont0=\footrm \scriptfont0=\footrms
\scriptscriptfont0=\footrmss
\textfont1=\footi \scriptfont1=\footis
\scriptscriptfont1=\footiss
\textfont2=\footsy \scriptfont2=\footsys
\scriptscriptfont2=\footsyss
\textfont\itfam=\footi \def\it{\fam\itfam\footi}
\textfont\mssfam=\footmss \def\mss{\fam\mssfam\footmss}
\textfont\bffam=\footbf \def\bf{\fam\bffam\footbf} \rm}
\def\tenpoint{\def\rm{\fam0\tenrm}
\textfont0=\tenrm \scriptfont0=\sevenrm
\scriptscriptfont0=\fiverm
\textfont1=\teni  \scriptfont1=\seveni
\scriptscriptfont1=\fivei
\textfont2=\tensy \scriptfont2=\sevensy
\scriptscriptfont2=\fivesy
\textfont\itfam=\tenit \def\it{\fam\itfam\tenit}
\textfont\mssfam=\tenmss \def\mss{\fam\mssfam\tenmss}
\textfont\bffam=\tenbf \def\bf{\fam\bffam\tenbf} \rm}
\ifx\answ\bigans\def\abstractfont{\tenpoint}\else
\def\abstractfont{\def\rm{\fam0\absrm}
\textfont0=\absrm \scriptfont0=\absrms
\scriptscriptfont0=\absrmss
\textfont1=\absi \scriptfont1=\absis
\scriptscriptfont1=\absiss
\textfont2=\abssy \scriptfont2=\abssys
\scriptscriptfont2=\abssyss
\textfont\itfam=\bigit \def\it{\fam\itfam\bigit}
\textfont\mssfam=\absmss \def\mss{\fam\mssfam\absmss}
\textfont\bffam=\absbf \def\bf{\fam\bffam\absbf}\rm}\fi
%
\def\f@@t{\baselineskip10pt\lineskip0pt\lineskiplimit0pt
\bgroup\aftergroup\@foot\let\next}
\setbox\strutbox=\hbox{\vrule height 8.pt depth 3.5pt width\z@}
\def\vfootnote#1{\insert\footins\bgroup
\baselineskip10pt\footfont
\interlinepenalty=\interfootnotelinepenalty
\floatingpenalty=20000
\splittopskip=\ht\strutbox \boxmaxdepth=\dp\strutbox
\leftskip=24pt \rightskip=\z@skip
\parindent=12pt \parfillskip=0pt plus 1fil
\spaceskip=\z@skip \xspaceskip=\z@skip
\Textindent{$#1$}\footstrut\futurelet\next\fo@t}
\def\Textindent#1{\noindent\llap{#1\enspace}\ignorespaces}
\def\foot{\global\advance\ftno by1%
\attach{\hyperref{}{footnote}{\the\ftno}{\footsymbolgen}}%
\vfootnote{\hyperdef\hypernoname{footnote}{\the\ftno}{\footsymbol}}}%
\def\footnote#1{\global\advance\ftno by1%
\attach{\hyperref{}{footnote}{\the\ftno}{#1}}%
\vfootnote{\hyperdef\hypernoname{footnote}{\the\ftno}{#1}}}%
\newcount\lastf@@t           \lastf@@t=-1
\newcount\footsymbolcount    \footsymbolcount=0
\global\newcount\ftno \global\ftno=0
\def\footsymbolgen{\relax\footsym
\global\lastf@@t=\pageno\footsymbol}
\def\footsym{\ifnum\footsymbolcount<0
\global\footsymbolcount=0\fi
{\iffrontpage \else \advance\lastf@@t by 1 \fi
\ifnum\lastf@@t<\pageno \global\footsymbolcount=0
\else \global\advance\footsymbolcount by 1 \fi }
\ifcase\footsymbolcount
\fd@f\dagger\or \fd@f\diamond\or \fd@f\ddagger\or
\fd@f\natural\or \fd@f\ast\or \fd@f\bullet\or
\fd@f\star\or \fd@f\nabla\else \fd@f\dagger
\global\footsymbolcount=0 \fi }
\def\fd@f#1{\xdef\footsymbol{#1}}
\def\space@ver#1{\let\@sf=\empty \ifmmode #1\else \ifhmode
\edef\@sf{\spacefactor=\the\spacefactor}
\unskip${}#1$\relax\fi\fi}
\def\attach#1{\space@ver{\strut^{\mkern 2mu #1}}\@sf}
%
\newif\ifnref
\def\rrr#1#2{\relax\ifnref\nref#1{#2}\else\ref#1{#2}\fi}
\def\ldf#1#2{\begingroup\obeylines
\gdef#1{\rrr{#1}{#2}}\endgroup\unskip}

\def\doubref#1#2{\refs{{#1},{#2}}}

\nreffalse
\def\refout{\footatend\vskip 2.cm\immediate\closeout\rfile\writestoppt
\baselineskip=\footskip\centerline{{\bf
References}}\bigskip{\parindent=20pt
\input \jobname.refs\vfill\eject}\nonfrenchspacing}
%
\def\eqn#1{\xdef #1{(\noexpand\hyperref{}%
{equation}{\secsym\the\meqno}%
{\secsym\the\meqno})}\eqno(\hyperdef\hypernoname{equation}%
{\secsym\the\meqno}{\secsym\the\meqno})\eqlabeL#1%
\writedef{#1\leftbracket#1}\global\advance\meqno by1}
\def\eqnalign#1{\xdef #1{\noexpand\hyperref{}{equation}%
{\secsym\the\meqno}{(\secsym\the\meqno)}}%
\writedef{#1\leftbracket#1}%
\hyperdef\hypernoname{equation}%
{\secsym\the\meqno}{\e@tf@ur#1}\eqlabeL{#1}%
\global\advance\meqno by1}
\def\eqnalign#1{\xdef #1{(\secsym\the\meqno)}
\writedef{#1\leftbracket#1}%
\global\advance\meqno by1 #1\eqlabeL{#1}}
%

%
\def\chap#1{\newsec{#1}}
\def\chapter#1{\chap{#1}}
\def\sect#1{\subsec{#1}}
\def\section#1{\sect{#1}}
\def\\{\ifnum\lastpenalty=-10000\relax
\else\hfil\penalty-10000\fi\ignorespaces}
\def\note#1{\leavevmode%
\edef\@@marginsf{\spacefactor=\the\spacefactor\relax}%
\ifdraft\strut\vadjust{%
\hbox to0pt{\hskip\hsize%
\ifx\answ\bigans\hskip.1in\else\hskip .1in\fi%
\vbox to0pt{\vskip-\dp
\strutbox\sevenbf\baselineskip=8pt plus 1pt minus 1pt%
\ifx\answ\bigans\hsize=.7in\else\hsize=.35in\fi%
\tolerance=5000 \hbadness=5000%
\leftskip=0pt \rightskip=0pt \everypar={}%
\raggedright\parskip=0pt \parindent=0pt%
\vskip-\ht\strutbox\noindent\strut#1\par%
\vss}\hss}}\fi\@@marginsf\kern-.01cm}
\def\titlepage{%
\frontpagetrue\nopagenumbers\abstractfont%
\hsize=\hstitle\rightline{\vbox{\baselineskip=10pt%
{\abstractfont\pubnum}}}\pageno=0}
\frontpagefalse
\def\pubnum{}
\def\pdate{\number\month/\number\yearltd}
\def\makefootline{\iffrontpage\vskip .27truein
\line{\the\footline}
\vskip -.1truein\leftline{\vbox{\baselineskip=10pt%
{\abstractfont\pdate}}}
\else\vskip.5cm\line{\hss \tenrm $-$ \folio\ $-$ \hss}\fi}
\def\title#1{\vskip .7truecm\titlestyle{\titleft #1}}
\def\titlestyle#1{\par\begingroup \interlinepenalty=9999
\leftskip=0.02\hsize plus 0.23\hsize minus 0.02\hsize
\rightskip=\leftskip \parfillskip=0pt
\hyphenpenalty=9000 \exhyphenpenalty=9000
\tolerance=9999 \pretolerance=9000
\spaceskip=0.333em \xspaceskip=0.5em
\noindent #1\par\endgroup }
\def\autskip{\ifx\answ\bigans\vskip.5truecm\else\vskip.1cm\fi}
\def\author#1{\vskip .7in \centerline{#1}}

\def\address#1{\ifx\answ\bigans\vskip.2truecm
\else\vskip.1cm\fi{\it \centerline{#1}}}
\def\abstract#1{
\vskip .5in\vfil\centerline
{\bf Abstract}\penalty1000
{{\smallskip\ifx\answ\bigans\leftskip 2pc \rightskip 2pc
\else\leftskip 5pc \rightskip 5pc\fi
\noindent\abstractfont \baselineskip=12pt
{#1} \smallskip}}
\penalty-1000}
\def\endpage{\tenpoint\supereject\global\hsize=\hsbody%
\frontpagefalse\footline={\hss\tenrm\folio\hss}}
\def\ack{\goodbreak\vskip2.cm\centerline{{\bf Acknowledgements}}}
%
%

%
\def\bfone{\relax{\rm 1\kern-.35em 1}}
\def\inbar{\vrule height1.5ex width.4pt depth0pt}
\def\IC{\relax\,\hbox{$\inbar\kern-.3em{\mss C}$}}
\def\ID{\relax{\rm I\kern-.18em D}}
\def\IF{\relax{\rm I\kern-.18em F}}
\def\IH{\relax{\rm I\kern-.18em H}}
\def\II{\relax{\rm I\kern-.17em I}}
\def\IN{\relax{\rm I\kern-.18em N}}
\def\IP{\relax{\rm I\kern-.18em P}}
\def\IQ{\relax\,\hbox{$\inbar\kern-.3em{\rm Q}$}}
\def\IR{\relax{\rm I\kern-.18em R}}
\font\cmss=cmss10 \font\cmsss=cmss10 at 7pt
\def\ZZ{\relax\ifmmode\mathchoice
{\hbox{\cmss Z\kern-.4em Z}}{\hbox{\cmss Z\kern-.4em Z}}
{\lower.9pt\hbox{\cmsss Z\kern-.4em Z}}
{\lower1.2pt\hbox{\cmsss Z\kern-.4em Z}}\else{\cmss Z\kern-.4em
Z}\fi}
\def\a{\alpha}  \def\d{\delta}

 \def\s{\sigma}

\def\nup#1({Nucl.\ Phys.\ $\us {B#1}$\ (}
\def\plt#1({Phys.\ Lett.\ $\us  {#1}$\ (}
\def\cmp#1({Comm.\ Math.\ Phys.\ $\us  {#1}$\ (}
\def\prp#1({Phys.\ Rep.\ $\us  {#1}$\ (}
\def\prl#1({Phys.\ Rev.\ Lett.\ $\us  {#1}$\ (}
\def\prv#1({Phys.\ Rev.\ $\us  {#1}$\ (}
\def\mpl#1({Mod.\ Phys.\ Let.\ $\us  {A#1}$\ (}
\def\ijmp#1({Int.\ J.\ Mod.\ Phys.\ $\us{A#1}$\ (}
\def\tit#1|{{\it #1},\ }
%

%

\def\ni{\noindent}
\def\tilde{\widetilde}
\def\bar{\overline}
\def\us#1{\underline{#1}}

\def\hat{\widehat}

\def\Coeff#1#2{{#1\over #2}}
\def\Coe#1.#2.{{#1\over #2}}
\def\coeff#1#2{\relax{\textstyle {#1 \over #2}}\displaystyle}
\def\coe#1.#2.{\relax{\textstyle {#1 \over #2}}\displaystyle}

\def\shalf{\relax{\textstyle {1 \over 2}}\displaystyle}

\def\to{\rightarrow}
\def\notin{\hbox{{$\in$}\kern-.51em\hbox{/}}}

\def\del{\partial}

\catcode`\@=12

\def\a{a_1}

\def\S{\Sigma}
\def\s{\sigma}

\def\wt{\tilde W}

\def\note{\vskip .3cm \ni$\spadesuit$\ }
\def\a{\sigma}
\def\o{{\cal O}}
\def\tauk{\tau_{K3}}

\def\eprt#1{{\tt #1}}
\def\nihil#1{{\sl #1}}
\def\br{\hfill\break}

\lref\witD{E.\ Witten,
 \nihil{Solutions of four-dimensional field theories via M theory,}
 \eprt{hep-th/9703166}.}

\lref\wlrev{W.\ Lerche,
 \nihil{Introduction to Seiberg-Witten theory and its stringy origin,}
 Nucl.\  Phys.\  Proc.\  Suppl.\ {\bf 55B} (1996) 83,
 \eprt{hep-th/9611190}.}
\lref\PM{P.\ Mayr,
 \nihil{Mirror symmetry, N=1 superpotentials and tensionles strings on
Calabi-Yau four folds,}
 Nucl.\  Phys.\ {\bf B494} (1996) 489,
 \eprt{hep-th/9610162}.}
\lref\KLRY{A.\ Klemm, B.\ Lian, S. Roan and S.\ Yau,
 \nihil{Calabi-Yau fourfolds for M theory and F theory
compactifications,}
 \eprt{hep-th/9701023}.}
\lref\akrev{A.\ Klemm,
 \nihil{On the geometry behind N=2 supersymmetric effective actions in
four-dimensions,}
 \eprt{hep-th/9705131}.}

\lref\witphases{E.\ Witten,
 \nihil{Phases of N=2 theories in two-dimensions,}
 Nucl.\  Phys.\ {\bf B403} (1993) 159-222,
 \eprt{hep-th/9301042}.}

\lref\grass{E.\ Witten,
 \nihil{The Verlinde algebra and the cohomology of the Grassmannian,}
 \eprt{hep-th/9312104}.}

\ldf\KLMVW{A.\ Klemm, W.\ Lerche, P.\ Mayr, C.\ Vafa and
N.\ Warner,
 \nihil{Selfdual strings and N=2 supersymmetric field theory,}
 Nucl.\  Phys.\ {\bf B477} (1996) 746-766,
 \eprt{hep-th/9604034}.}
\lref\KKV{S.\  Katz, A.\  Klemm and C.\ Vafa,
 \nihil{Geometric engineering of quantum field theories,}
 Nucl.\ Phys.\ {\bf B497} (1997) 173-195,
 \eprt{hep-th/9609239}.}

\lref\KMV{S.\ Katz, P.\ Mayr and C.\ Vafa,
 \nihil{Mirror symmetry and exact solution of 4-D N=2 gauge theories:
I}
 \eprt{hep-th/9706110}; part II, to appear.}

\lref\KVsuppot{S.\ Katz and C.\ Vafa, first reference in
\geometricengineering.}

\lref\geometricengineering{S.\ Katz and C.\ Vafa,
 \nihil{Geometric engineering of N=1 quantum field theories,}
Nucl.\ Phys.\ {\bf B497} (1997) 196-204,
 \eprt{hep-th/9611090}; \br
  {M.\ Bershadsky, A.\ Johansen, T.\ Pantev, V.\ Sadov and C.\ Vafa,
 \nihil{F theory, geometric engineering and N=1 dualities,}
 \eprt{hep-th/9612052}; \br}
{H.\  Ooguri and C.\ Vafa,
 \nihil{Geometry of N=1 dualities in four-dimensions,}
 \eprt{hep-th/9702180}.}
}
\lref\electricalengineering{See for example:\br {S.\ Elitzur, A.\
Giveon and D.\ Kutasov, \nihil{Branes and N=1 duality in string
theory,} Phys.\  Lett.\ {\bf B400} (1997) 269-274,
\eprt{hep-th/9702014}; \br} {Kentaro Hori, H.\ Ooguri and Yaron Oz,
\nihil{Strong coupling dynamics of four-dimensional N=1 gauge
theories from M theory five-brane,} \eprt{hep-th/9706082}.} }

\lref\MQCD{E.\ Witten,
 \nihil{Branes and the dynamics of QCD,}
 \eprt{hep-th/9706109}.}

\lref\KVa{S.\  Katz and C.\ Vafa,
 \nihil{Matter from geometry,} Nucl.\ Phys.\ {\bf B497} (1997) 146-154,
 \eprt{hep-th/9606086}.}

\lref\delP{ {W.\ Lerche, P.\ Mayr and N.\ Warner, \nihil{Noncritical
strings, Del Pezzo singularities and Seiberg-Witten curves,}  Nucl.\
Phys.\ {\bf B499} (1997) 125, \eprt{hep-th/9612085}.} }

\lref\QV{B.\ Greene and Y.\ Kanter,
 \nihil{Small volumes in compactified string theory,}
 Nucl.\ Phys.\ { \bf B497} (1997) 127-145,
 \eprt{hep-th/9612181}.}

\lref\HW{A.\  Hanany and E.\ Witten,
 \nihil{Type IIB superstrings, BPS monopoles, and three-dimensional
gauge dynamics,}
 Nucl.\  Phys.\ {\bf B492} (1997) 152-190,
 \eprt{hep-th/9611230}.}
\lref\LW{W.\ Lerche and N.\ P.\ Warner,
 \nihil{Exceptional SW geometry from ALE fibrations,}
 \eprt{hep-th/9608183}.}
\lref\KP{S.\ Gates, Jr., C.\ Hull and M.\ Rocek,
 \nihil{Twisted multiplets and new supersymmetric nonlinear sigma
models,}
 Nucl.\  Phys.\ {\bf B248} (1984) 157.}
\lref\FMS{S.\ Ferrara, R.\ Minasian and A.\ Sagnotti,
 \nihil{Low-energy analysis of M and F theories on Calabi-Yau
threefolds,}
 Nucl.\  Phys.\ {\bf B474} (1996) 323-342,
 \eprt{hep-th/9604097}.}
\lref\colem{S.\ Coleman, \nihil{More on the massive Schwinger model,}
Ann.\ Phys.\ {\bf 101} (1976) 239.}
\lref\Fth{{C.\ Vafa,
 \nihil{Evidence for F theory,}
 Nucl.\  Phys.\ {\bf B469} (1996) 403-418,
 \eprt{hep-th/9602022}; \br}
D.\  Morrison and C.\ Vafa,
 \nihil{Compactifications of F theory on Calabi-Yau threefolds I,}
 Nucl.\  Phys.\ {\bf B473} (1996) 74-92,
 \eprt{hep-th/9602114}; \br{
 \nihil{Compactifications of F theory on Calabi-Yau threefolds II,}
 Nucl.\  Phys.\ {\bf B476} (1996) 437-469,
 \eprt{hep-th/9603161}.}
}
\lref\stro{J.\ Polchinski and A.\ Strominger,
 \nihil{New vacua for type II string theory,}
 Phys.\  Lett.\ {\bf B388} (1996) 736-742,
 \eprt{hep-th/9510227}.}
\lref\witflux{E.\ Witten,
 \nihil{On flux quantization in M theory and the effective action,}
 \eprt{hep-th/9609122}.}
\lref\SVW{S.\ Sethi, C.\ Vafa and E.\ Witten,
 \nihil{Constraints on low-dimensional string compactifications,}
 Nucl.\  Phys.\ {\bf B480} (1996) 213-224,
 \eprt{hep-th/9606122}.}
\lref\GMP{B.\ Greene, D.\ Morrison and M.\ Plesser,
 \nihil{Mirror manifolds in higher dimension,}
 Commun.\  Math.\  Phys.\ {\bf 173} (1995) 559-598,
 \eprt{hep-th/9402119}.}
\lref\LScW{W.\ Lerche, A.\ Schellekens and N.\ Warner,
 \nihil{Lattices and strings,}
 Phys.\  Rept.\ {\bf 177} (1989) 1.}
\lref\SV{S.\ Shatashvili and C.\ Vafa,
 \nihil{Superstrings and manifold of exceptional holonomy,}
 \eprt{hep-th/9407025}.}
\lref\KKLMV{ShA.\  Kachru, A.\  Klemm, W.\ Lerche, P.\  Mayr and C.\
Vafa,  \nihil{Nonperturbative results on the point particle limit of
N=2 heterotic string compactifications,}  Nucl.\  Phys.\ {\bf B459}
(1996) 537-558,  \eprt{hep-th/9508155}.}
\lref\LSW{W.\ Lerche, D.\ Smit and N.\ Warner,  \nihil{Differential
equations for periods and flat coordinates in two-dimensional
topological matter theories,}  Nucl.\  Phys.\ {\bf B372} (1992) 87-112,
 \eprt{hep-th/9108013}.}

 \lref\LY{B.\ Lian and S.\ Yau,  \nihil{Arithmetic properties of mirror
map and quantum coupling,}  Commun.\  Math.\  Phys.\ {\bf 176} (1996)
163-192,  \eprt{hep-th/9411234}; \nihil{Mirror maps, modular relations
and hypergeometric series 1,2}
\eprt{hep-th/9507151}, \eprt{hep-th/9507153}. }
\lref\VW{C.\ Vafa and E.\ Witten,
 \nihil{A One loop test of string duality,}
 Nucl.\  Phys.\ {\bf B447} (1995) 261-270,
 \eprt{hep-th/9505053}.}

\lref\Jan{Jan de Boer, private communication.}

\lref\spec{L.\ Dixon, V.\ Kaplunovsky and J.\ Louis,  \nihil{On
effective field theories describing (2,2) vacua of the heterotic
string,}  Nucl.\  Phys.\ {\bf B329} (1990) 27-82; \br {M.\ Bershadsky,
S.\ Cecotti, H.\ Ooguri and C.\ Vafa,  \nihil{Kodaira-Spencer theory of
gravity and exact results for quantum string amplitudes,}  Commun.\
Math.\  Phys.\ {\bf 165} (1994) 311-428,  \eprt{hep-th/9309140}.}
}
\lref\beckerSquared{K.\ \&\ M.\ Becker,
 \nihil{M theory on eight manifolds,}
 Nucl.\  Phys.\ {\bf B477} (1996) 155-167,
 \eprt{hep-th/9605053}.}
\lref\philip{P.\ Candelas, X.\ De La Ossa, P.\ Green and
L.\ Parkes,  \nihil{A Pair of Calabi-Yau manifolds as an exactly
soluble superconformal theory,}  Nucl.\  Phys.\ {\bf B359} (1991)
21-74.}

\lref\MP{D.\  Morrison and M.\ Ronen Plesser,
 \nihil{Summing the instantons:
 Quantum cohomology and mirror symmetry in toric varieties,}
 Nucl.\  Phys.\ {\bf B440} (1995) 279-354,
 \eprt{hep-th/9412236}.}

\lref\DM{K.\ Dasgupta and S.\ Mukhi,
 \nihil{A Note on low dimensional string compactifications,}
 Phys.\  Lett.\ {\bf B398} (1997) 285-290,
 \eprt{hep-th/9612188}.}

\lref\EMF{See eg.: B.\ Sch\"oneberg, {\it Elliptic Modular Functions},
Springer 1974, page 95/96.}

\lref\nagu{M.\ Nagura and K.\ Sugiyama,
 \nihil{Mirror symmetry of K3 and surface,}
 Int.\  J.\  Mod.\  Phys.\ {\bf A10} (1995) 233-252,
 \eprt{hep-th/9312159}.}

\lref\SW{N.\ Seiberg and E.\ Witten, \nup426(1994) 19,
\eprt{hep-th/9407087}; \nup431(1994) 484, \eprt{hep-th/9408099}.}

\lref\HH{A.\ Hanany and K.\ Hori,
 \nihil{Branes and N=2 theories in two-dimensions,}
 \eprt{hep-th/9707192}.}

\lref\HS{{J.\ Harvey and A.\ Strominger,
 \nihil{The heterotic string is a soliton,}
 Nucl.\  Phys.\ {\bf B449} (1995) 535-552,
 \eprt{hep-th/9504047}.}
}

\lref\CAF{A.\
Ceresole, R.\ D'Auria and S.\ Ferrara, \nihil{On the Geometry of Moduli
Space of Vacua in $N=2$ Supersymmetric Yang-Mills Theory,} Phys.\
Lett.\ {\bf B339} (1994) 71-76, hep-th/9408036;}

\lref\Z{{E.\ Witten and D.\ Olive,  \nihil{Supersymmetry algebras that
include topological charges,}  Phys.\  Lett.\ {\bf 78B} (1978) 97.};\br
{P.\ Fendley, S.\ Mathur, C.\ Vafa and N.\ Warner,  \nihil{Integrable
deformations and scattering matrices for the N=2 supersymmetric
discrete series,}  Phys.\  Lett.\ {\bf B243} (1990) 257-264.};\br {S.\
Cecotti and C.\ Vafa,  \nihil{On classification of N=2 supersymmetric
theories,}  Commun.\  Math.\  Phys.\ {\bf 158} (1993) 569-644,
\eprt{hep-th/9211097}.} }

 \lref\AHISS{{O.\ Aharony, A.\ Hanany, K.\ Intriligator,  N.\ Seiberg
and M.\ Strassler,  \nihil{Aspects of N=2 supersymmetric gauge theories
in three dimensions,}  Nucl.\  Phys.\ {\bf B499} (1997) 67,
\eprt{hep-th/9703110}.} }

\lref\KKLMV{S.\ Kachru, A.\ Klemm, W.\ Lerche, P.\ Mayr and  C.\ Vafa,
\nihil{Nonperturbative results on the point particle limit of N=2
heterotic string compactifications,}  Nucl.~ Phys.~{\bf B459} (1996)
537-558,  \eprt{hep-th/9508155}.}

\lref\GSW{M.\ Green, J.\ Schwarz and E.\ Witten, {\it Superstring
Theory}, Vol.\ 2, Cambridge University Press 1987, page 456.}

\lref\superpot{
{E.\ Witten,
 \nihil{Nonperturbative superpotentials in string theory,}
 Nucl.\  Phys.\ {\bf B474} (1996) 343-360,
 \eprt{hep-th/9604030}.}
}

\lref\DMW{
{M.\ Duff, R.\ Minasian and E.\ Witten,
 \nihil{Evidence for heterotic / heterotic duality,}
 Nucl.~ Phys.~{\bf B465} (1996) 413-438,
 \eprt{hep-th/9601036}.}
}

\def\pubnum{
\hbox{CERN-TH/97-247}
\hbox{hep-th/9709146}}
\def\pdate{}
\titlepage
\vskip2.cm
\title
{{\titlefont Fayet-Iliopoulos Potentials from Four-Folds}}
\autskip
\author{W.\ Lerche}
\vskip0.2truecm
\CERN
\vskip-1.2truecm

\abstract
{
We show how certain non-perturbative superpotentials $\tilde
W(\Sigma)$, which are the two-dimensional analogs of the Seiberg-Witten
prepotential in 4d, can be computed via geometric engineering from
4-folds. We analyze an explicit example for which the relevant compact
geometry of the 4-fold is given by $\IP^1$ fibered  over $\IP^2$. In
the field theory limit, this gives an effective $U(1)$ gauge theory
with $N=(2,2)$ supersymmetry in two dimensions. We find that the analog
of the SW curve is a $K3$ surface, and that the complex FI coupling is
given by the modular parameter of this surface. The FI potential itself
coincides with the middle period of a meromorphic differential.
However, it only shows up in the effective action if a certain 4-flux
is switched on, and then supersymmetry appears to be non-perturbatively
broken. This can be avoided by tuning the bare FI coupling by hand, in
which case the supersymmetric minimum naturally corresponds to a
singular $K3$.
}

\vfil
\vskip 1.cm
\ni {CERN-TH/97-247}\hfill\break
\ni September  1997
\endpage
\baselineskip=14pt plus 2pt minus 1pt


\chapter{Introduction}

String duality has proven to be an extremely useful tool for
investigating non-perturbative properties of supersymmetric gauge and
other types of theories.  Currently there are two
complementary approaches: ``geometric engineering"
\refs{\KLMVW{,}\KVa{,}\LW{,}\KKV{,}\geometricengineering{,}\KMV}\ makes
use of the local singular geometry of compactification manifolds, while
the other approach, originating in \HW, uses
(essentially) parallel flat $D$-branes to model the relevant geometry.
The relation between these approaches has recently been illuminated in
\witD\ for $N=2$ gauge theories in four dimensions.

The power of the geometric approach for solving gauge theories is that
all gauge groups can be treated systematically in the same way. So far,
most of the more concrete results that have been obtained from
geometric engineering \refs{\KKLMV,\KLMVW,\KKV,\KMV}\ concern $N=2$
supersymmetric gauge theories in
$d=4$ \SW\foot {For reviews see \doubref\wlrev\akrev.}, but
obviously theories with $N=1$ supersymmetry are phenomenologically more
important. Such theories have been investigated in
\geometricengineering, and from the $D$-brane point of view, for
example in refs.\ \doubref\electricalengineering\MQCD.

It would certainly be interesting to apply similar methods to obtain
non-trivial information about $N=1$ supersymmetric theories in four
dimensions. Since such theories can be obtained from $F$-theory
compactifications on elliptic 4-folds $X$ \Fth, this suggests to
study ``local" (or ``rigid") mirror symmetry\foot{General aspects of
mirror symmetry of d-folds have been first discussed in \GMP;
specifically 4-folds were analyzed in detail in \PM\ and subsequently
in \KLRY.} of the relevant singular geometries of these 4-folds.
However, $F$-theory is at present still quite hard to deal with
directly, and it seems a bit simpler to study instead
compactifications of type IIA strings on the same kind of 4-folds.
Such compactifications lead to $N=(2,2)$ supersymmetric theories\foot
{In in D-brane language, such theories have recently been
investigated in \HH.} in $d=2$, which may be viewed as reductions of
the corresponding $N=1$ theories in four dimensions. In type II
string theory we can then use mirror symmetry in a more
straightforward fashion, but nevertheless may expect that some of the
relevant novel features of 4-folds can be captured in this simplified
two-dimensional setting. It is the purpose of the present paper to
gain some insight in how 4-folds work, by investigating a specific
example.

\chapter{Holomorphic Fayet-Iliopoulos potentials}

Among of the most basic problems  are 4-fold geometries that lead to
gauge theories in two dimensions. Certainly gauge fields do not
propagate in $d=2$, but the scalar components $\s$ of the $N=(2,2)$
supermultiplets do. More specifically, the relevant supermultiplets are
the ``twisted chiral'' field strength multiplets,
$\S\sim\sigma+\dots+\theta^+\bar\theta^-(D-iF)$, which obey $\bar
D_+\S=D_-\S=0$, while matter fields correspond to the ordinary chiral
multiplets $\Phi$ with $\bar D_\pm\Phi=0$, etc. The most general
lagrangian involving these two sorts of fields \KP\ consists of a
generalized K\"ahler potential $K(\S,\bar\S,\Phi,\bar\Phi)$ plus
holomorphic chiral and twisted chiral potentials, $W(\Phi)$ and $\tilde
W(\S)$. In absence of chiral matter multiplets $\Phi$, the twisted
chiral multiplets $\S$ are equivalent to the $\Phi$, and can be
transformed into them \KP.

Thus, integrating out massive chiral matter fields, the effective
action of a gauge theory will simply be a twisted $(2,2)$
supersymmetric sigma-model, given by some $K(\S,\bar\S)$ plus possibly
some twisted chiral potential $\wt(\S)$. We take the scaling dimension
of $\Sigma$ to be equal to one, so that the K\"ahler potential has to
be multiplied by the squared inverse of a dimensionful gauge coupling
and so becomes an irrelevant operator in the infrared. The other piece
of the lagrangian, the twisted chiral potential, plays the r\^ole of a
generalized Fayet-Iliopoulos term \refs{\witphases,\grass,\MP,\HH}:
$$
{i\over2\sqrt2}\int d\theta^+d\bar\theta^-\, \wt(\S)\ + {\rm c.c.}\ =\
-\xi(\s)\,D+{\theta(\s)\over2\pi}\,F\ .
\eqn\WDterm
$$
It gives rise to an effective, field dependent complex FI
coupling:
$$
\tau(\s)\ \equiv\  i\,\xi(\s)+{\theta(\s)\over2\pi}\ =\ \wt'(\s)\ .
\eqn\taueff
$$
The twisted chiral potential $\wt(\s)$ is the semi-topological,
holomorphic quantity that is the analog of the SW prepotential ${\cal
F}(a)$ \SW\ in four dimensions, and which is of our main interest. The
FI coupling $\tau$ is the analog of the running gauge coupling in 4d,
being dimensionless and subject to RG flow. Indeed, it is known
\witphases\ that $\tau$ receives logarithmic perturbative
corrections to exactly one loop order,
$$
\tau(\s)=\tau_0- {N\over 2\pi i}\log(\s/\mu)+\dots...\ ,
\eqn\tauren
$$
where $\mu$ is the RG scale. If we interpret the sigma-model in terms
of a gauge theory, then $N=TrQ$, where $Q$ is the $U(1)$ charge
of the charged chiral matter fields.

Clearly, logarithmic monodromy shifts induce shifts of the
theta-angle, exactly like for $N=2$ gauge theories in four
dimensions. For positive $N$ and Im$\tau$, the theory is
asymptotically free in the FI coupling, which means weakly coupled
for large $\s$. We generically expect additional non-perturbative
corrections to $\tau(\s)$ in \tauren, the $n$-th instanton sector
being weighted by $\beta^n$, where $\beta=e^{2\pi i\tau_0}\mu^N\equiv
e^{i\theta_0-2\pi \xi_0}\mu^N$.

An important difference as compared to the four dimensional
gauge theory is that there is a non-trivial scalar potential:
$$
V(\s)\ =\ {1\over2} |\tau(\s)|^2\ .
\eqn\Vpot
$$
This means in particular that the vacuum energy depends on the
theta-angle \doubref\colem\witphases. It also means that supersymmetry
is broken if $\tau(\s)$ is everywhere non-vanishing. Semi-classically,
where we only consider the perturbative correction in \tauren, there
will be $N$ vacuum states, given by $\s=\beta^{1/N}$ plus rotations by
the $\ZZ_{2N}$ $R$-symmetry; each VEV breaks the $R$-symmetry to
$\ZZ_2$.

\chapter{Mirror Symmetry}

Like in four dimensions, the gauge multiplets $\S$ are one-to-one
to the K\"ahler classes belonging to $H^{1,1}(X)$, while the chiral
matter fields correspond to the complex structure moduli belonging to
$H^{3,1}(X)$. Since the type IIA dilaton is in a gravitational
multiplet, which is real rather than twisted chiral, the holomorphic
twisted chiral potential $\wt(\S)$ does not get any type IIA
space-time corrections. On the other hand, there will in general be
corrections from world-sheet instantons to the K\"ahler sector,
reflecting perturbative and non-perturbative corrections in
the dual heterotic string language.  The issue is to compute these
corrections to $\wt(\S)$ via mirror symmetry, which maps the type IIA
string on the 4-fold $X$ back to the type IIA string on the mirror
4-fold, $\hat X$. In addition, the r\^oles of complex structure and
K\"ahler moduli get exchanged. Thus in the mirror theory
the complex structure sector is not corrected at all,
as there are no 3-branes in the type IIA string that could wrap the
middle homology 4-cycles, and a tree-level computation is exact.

In order to see what precise tree-level correlator we will
have to compute to obtain $\wt(\S)$, consider the following tree-level
Chern-Simons term in the $d=10$ type IIA string:
$$
{\cal L}_{CS}\ =\ B\wedge F_4\wedge F_4\ ,
\eqn\LCS
$$
where $F_4$ is the field strength of the 3-form field.
On the 4-fold $X$ we can then expand
$$
B\ =\ \s\, \o^{(1)}\ ,\qquad\ \ {F_4}\ =\ \nu\, \o^{(2)} +
F\wedge \o^{(1)}\ ,
\eqn\formexpand
$$
where $\o^{(i)}$ represent elements of $H^{i,i}_{\bar\del}(X,\ZZ)$. The
occurrence of 4-forms $\o^{(2)}$ \doubref\PM\KLRY\ and the related
scalars $\nu$ is a novel feature as compared to usual 3-fold
compactifications. The important point is that at the quantum level,
$\nu$ is an integer c-number:
$$
\int_{{\cal C}^4}F_4\ =\nu\ \in\ZZ
\eqn\fourflux
$$ (or possibly $\nu \in{1\over2}\ZZ$),
which is known as ``4-flux'' \refs{\HS{,}\stro{,}\witflux}. Therefore
\LCS\ leads to a two-dimensional FI coupling of the form
$\s\,F\,\nu\,\langle\o^{(1)}\o^{(1)}\o^{(2)}\rangle$, which means that
$$
\wt''(\s)\ =\
\nu\,\langle\,\o^{(1)}\,\o^{(1)}\,\o^{(2)}\,\rangle_{{IIA}}
\ .\eqn\wdpr
$$
We see that the FI coupling is proportional to the 4-flux, and since
non-trivial FI terms typically do exist, $\nu$ will generically have to
be non-zero. On the other hand, unbroken supersymmetry tends to favor
$\nu=0$, though this is not strictly required
\refs{\stro{,}\beckerSquared}. We will come back to this
point later in section 5.2.

One might wonder about fivebrane instantons wrapping around appropriate
6-cycle divisors, producing corrections to the K\"ahler sector that may
not be captured by mirror symmetry; indeed such instantons do lead to
potentials in $M$- and $F$-theory compactifications to three and four
dimensions \doubref\superpot\KVsuppot.  There is a simple argument why
such instantons do not contribute to $\tilde W$, the reason being the
non-zero flux $\nu$.   More specifically, it is known that on the
fivebrane world volume, there is a self-dual 2-form field with field
strength $T_3$ which obeys $dT_3=F_4$. As pointed out in \DMW, from
this follows that a wrapped 5-brane implies $F_4$ to be cohomologically
trivial, ie., $\nu=0$ in \fourflux. Conversely, a non-zero 4-flux
(emanating from a submanifold ${\cal C}^4$ of the 6-cycle) prohibits
the wrapping of the fivebrane, and thus there are no fivebrane
instanton corrections to $\tilde W$.\foot{This makes sense also from
the point of view of $M$- and $F$-theory compactifications on $X$,
where
$\nu \langle\,\o^{(1)}\,\o^{(1)}\,\o^{(2)}\,\rangle_{{\rm class}}$  is
the coefficient of Chern-Simons and GS anomaly cancelling terms,
respectively. Such topological couplings are supposed not to be
corrected.}

The 3-point function in \wdpr\ is a correlator in topological field
theory
that can be evaluated via mirror symmetry \refs{\GMP{,}\PM{,}\KLRY}.
When twisting by the internal $U(1)$ current we can project on either
the chiral or on the twisted chiral subsector of the theory. When
twisting left-right symmetrically, the background charges are
$(-4,-4)$ and thus we project on the $\S$-sector where the basic
correlators are
$$
C_{112}\ =\ \langle\,\o^{(1)}\,\o^{(1)}\,\o^{(2)}\,\rangle\ ,
\qquad C_{1111}\ =\
\langle\,\o^{(1)}\,\o^{(1)}\,\o^{(1)}\o^{(1)}\,\rangle
\ .\eqn\corrs
$$
Both $\o^{(1)}$ and $\o^{(2)}$ survive the topological twist (as
they obey $h=q/2$), even though the 4-form operators do not represent
continuous but discrete moduli of the TFT. Note that only the
three-point function and not the four-point function contributes to a
holomorphic potential. Indeed after twisting (and setting $\o^{(i)}\to
\o^{(i)}e^{-\phi-\bar\phi}$), one needs exactly three and not four
operators in the $(-1,-1)$-picture, and therefore one has to insert an
extra picture changing operator in the four-point correlator. This
introduces momentum factors which means that the four-point function
contributes to the K\"ahler potential and not to $\tilde W$.

Note also that on 3-folds, the 4-forms are dual to 2-forms so that
they are not independent. In contrast, for 4-folds the $H^{2,2}$
sector is independent and generically quite large, and in general
only a small subset of $H^{2,2}$ will be related to the sub-sector we
are interested in (the "primary subspace" generated by wedging the
(1,1)-forms). In particular, for theories with one modulus, the
relevant 4-form is simply $\o^{(2)}\sim (\o^{(1)})^2$, and therefore
we have from factorization \GMP:
$$
{C_{1111} \ \sim\ \big(C_{112}\big)^2\ .
}\eqn\factor
$$

\chapter{An example: $\IP^2$ fibered over $\IP^1$}

The best understood example for geometric engineering  is pure $N=2$
Yang-Mills theory \SW\ in $d=4$, where the relevant type IIA 3-fold
geometry is given by an $A_1$ singularity fibered over $\IP^1$ \KLMVW.
Since such a singularity describes a vanishing 2-sphere, the local
geometry is effectively given by a fibration of $\IP^1$ over $\IP^1$,
ie., by a Hirzebruch surface.\foot {Together with the normal bundle on
it, this yields a non-compact 3-fold whose compact part is
the fibration. We will in the following not explicitly mention the
non-compact parts of 3- or 4-folds.} The type IIB mirror geometry
of this fibration is indeed exactly given by (a non-compact form of)
the SW curve \doubref\KKV\KMV.

Our intention is to stay as close as possible to this situation, and
simply to try to see what will come out for a 4-fold as compared to a
3-fold. The closest relative of the SW geometry for a 4-fold is given
by a fibration of the same $\IP^1$ over a $\IP^2$ base. In the field
theory limit, one naively expects this to give rise to a reduction on
$\IP^2$ of a six-dimensional $SU(2)$ gauge theory down to two
dimensions, the $SU(2)$ arising from the fiber $\IP^1$. At any rate,
whether this expectation bears out or not, the low-energy effective
theory that we will obtain is a $U(1)$ gauge theory, and our task is to
compute its twisted chiral potential $\tilde W$.

On general grounds \doubref\PM\KLRY, the contributions from world-sheet
instantons (including multi-covers) to the potential will be of the
form  $\tilde W= Q(t_f,t_b)+\sum n_{i,j} {\rm Li}_2(q_f^i q_b^j)$
(where $Q$ is a quadratic function of the K\"ahler moduli associated
with fiber and base, and $q_{f,b}\equiv e^{2\pi i t_{f,b}}$).  It will
turn
out, exactly like in four dimensions \KKV, that wrappings of
world-sheet instantons around the $\IP^1$ fiber produce, in the rigid
limit, the logarithmic one-loop contribution to $\tilde W$ (from
Li$_2(1+\sqrt{\alpha'}\sigma)\sim {\rm const}+ \sqrt{\alpha'}\s{\rm
log}\s$), and wrappings around the various classes of the $\IP^2$ base
give additional non-perturbative corrections.

\subsec{Picard-Fuchs system}

Taking toric geometry as starting point, we choose
the following Mori (charge) vectors to describe the fibration data of
the non-compact 4-fold:
$$
\eqalign{
l_f\ &=\ (-2,1,0,0,0,1)\ ,\qquad (\IP^1 \ {\rm fiber})
 ,\cr
l_b\ &=\ (0,0,1,1,1,-3)\ ,\qquad (\IP^2 \ {\rm base})\ .\cr
}\eqn\mori
$$
Following standard methods (see eg., \akrev),
the non-compact mirror then looks, up to quadratic pieces:
$${
W\ =\
{z_b}\,{{x_2}^4} +
  {z_f}\,{{x_1}^2}\,x_3\,x_4 +
  x_1\,x_2\,x_3\,x_4 + {{x_2}^2}\,x_3\,x_4 +
  x_2\,{{x_3}^2}\,x_4 + x_2\,x_3\,{{x_4}^2}
}\eqn\Kt
$$
(which involves the canonical parameters:
$z_f = {a_2 a_6\over a_1^2}$, $z_b = {a_3 a_4 a_5\over a_6^3}$).
The associated Picard-Fuchs system is ($\theta_f\equiv z_f\del_{z_f}$,
etc):
$$
\eqalign{
L_f\ &=\ \theta_f(\theta_f-3 \theta_b)-2 z_f \theta_f (2\theta_f+1)\cr
L_b\ &=\ {\theta_b}^3- z_b (\theta_f-3 \theta_b-2)(\theta_f-3
\theta_b-1)
(\theta_f-3 \theta_b)\ ,\cr
}$$
and we find for the relevant components of the discriminant:
$$
\eqalign{
\Delta_1\ &=\ (1+27 z_b)^4\cr
\Delta_2\ &=\ (1-4 z_f)^3- 1728 z_f^3 z_b\ .\cr
}$$
Note the cubic splitting of the classical $A_1$ singularity at
$z_f=1/4$. The rigid field theory limit we are interested in, amounts
to taking the base $\IP^2$ large and the fiber $\IP^1$ small. This is
localized in the moduli space at the point of tangency $z_f=1/4,\,
z_b=0$, which needs to be properly blown-up \KKLMV. Suitable variables
for this double-scaling limit are given by:
$$
\eqalign{
z_1\ =\ 4 z_f-1\ \equiv\ \alpha' u\cr
z_2\ =\ -{3 {z_b}^{1/3}\over 4 z_f-1}\ \equiv\ {\beta^{1/3}\over u}\ .
}$$
which leads to $\Delta_2\sim{\alpha'}^3 (u^3-\beta)+O({\alpha'}^4)$.
Hence we are left with only one independent variable in the rigid
limit $\alpha'\to0$, so that effectively one $U(1)$ factor
decouples. We also see that by dimensional transmutation a scale $\mu$
is introduced in this process, $z_b\sim e^{-S}\sim(\alpha')^3\mu^6
e^{2\pi i\tau_0}$, where $S$ is the heterotic dilaton. This means that
$\beta\equiv e^{2\pi i\tau_0}\mu^6$, and that $u$ has mass dimension 2.
{}From now on, we will mostly set $\beta=1$.

After rescaling the periods, the above PF system reduces to the
following differential operator:
$$
\eqalign{
L_f\ &\to\ 0\cr
L_b\ &\to\ L_R\ \equiv\ (2\theta_2+1)^3- 8 {z_2}^3
\theta_2(\theta_2+1)(\theta_2+2)\ .\cr
}$$
Note that this operator, coming from the base $\IP^2$
and not from the fiber, is of third order, which is the natural
order of a PF operator associated with a 2-fold.
In terms of the inverse variable $u=1/z_2$, it constitutes
a generalized hypergeometric equation of type\foot
{For $SU(2)$ SW theory in d=4, the analogous system is
of type ${}_2F_1(-\coeff14,-\coeff14;\coeff12;u^2)$.}
${}_3F_2(-\coeff16,-\coeff16,-\coeff16;\coeff13,\coeff23;u^3)$, with
standard series solutions around $u=0$.

The moduli space thus has three singularities at $u^3=1$, plus one at
$u=\infty$, which is the weak-coupling limit of the rigid theory. In
the weak-coupling region, the three solutions behave as $\s\sim \sqrt
u$, $\tilde\s_{D1}\sim \sqrt u\log u$,  $\tilde\s_{D2}\sim \sqrt u(\log
u)^2$, where $\s$ represents the scalar component of the $U(1)$ gauge
superfield $\Sigma$. Since $u$ has mass dimension two, $\sigma$ has
mass dimension equal to one, and this exactly what we had assumed in
section 2.

\ni Moreover, we find that the derivatives
$$
{{\del\over\del u}\pmatrix {\s\cr \s_{D1} \cr \s_{D2}}(u)
\ \equiv\ \pmatrix{\omega\cr \omega_{D1} \cr
\omega_{D2}}(u)}\eqn\Kper
$$
are solutions of a generalized hypergeometric equation of type
${}_3F_2(\coeff16,\coeff16,\coeff16;\coeff13, \coeff23;u^3)$. These
correspond to the periods of the holomorphic 2-form,
$
\Omega_{2,0} \equiv ({\del W_{K3}\over\del x}
)^{-1}{dz\over z} \wedge  {dw\over w}
$,
that is associated with the following ``rigid'' $K3$ surface:
$$
W_{K3}\ =\ z+w-{1\over27zw}+(x^2+u)\ =\ 0\ .
\eqn\rigidKthree
$$
This $K3$, which arises from \Kt\ in the rigid limit after appropriate
rescalings, is the 2d analog of the elliptic curve \SW\ in four
dimensions.
Indeed from \Kper\ we
conclude, in complete analogy to SW theory, that
$\s, \s_{D1}, \s_{D2}$ are periods of a specific meromorphic 2-form
$$
\lambda_2\ =\ -2\,x\,{dz\over z} \wedge  {dw\over w}\ =\
-2i\sqrt{z+w-{1\over27zw}+u}\ {dz\over z} \wedge  {dw\over w}
$$
on the auxiliary $K3$ surface, which has
the characteristic property
$$
{\del\over\del u}\lambda_2\ =\ \Omega_{2,0}\ .
\eqn\derel
$$

\subsec{Properties of the periods}

Note that $K3$ periods are
algebraically dependent, $\omega \omega_{D2} \sim {\omega_{D1}}^2$,
whence there is only one independent ratio:
$$
{\tauk(u)\ =\ {\del_u \a_{D1}(u)\over \del_u \a(u) }
\ ,\ \ \ \ \
{\del_u \a_{D2}(u)\over \del_u \a(u) }\ =\ (\tauk(u))^2.
}\eqn\taudef
$$
In fact, it is known for some while \doubref\LSW\LY\ that the PF
equation of a single-modulus $K3$ can be reduced to a Schwarzian
differential equation for $\tau$. Specifically, following \LSW\ we find
from the $K3$ PF equation:
$$
\big\{\tauk(u); u^3 \big\} =\
\Coeff{36 {u^6}-41 {u^3}+32}{72 {u^6} {{({u^3}-1)}^2}}\ ,
\eqn\swar
$$
which implies that
$$
u^3(\tauk)\ =\ {1\over 1728}\,j\,(\tauk) \ ,
\eqn\taurel
$$
where $j$ is the well-known modular invariant.
Hence the monodromy group generated in the $u^3$-plane (with
singularities at $0,1,\infty$) acting on $\tauk$ is the modular group
$SL(2,\ZZ)$; this is indeed a typical feature of $K3$ surfaces \LY. The
monodromy group for our preferred parametrization, which is the
$u$-plane, is then the corresponding index 3 normal subgroup $\Gamma_1$
\EMF\ with branch scheme $(1,3,3)$ (with $T^3:\tauk\to\tauk+3$ as one
of its generators).

The specific linear combinations of the PF solutions that correspond
to the integral geometric periods can be determined in various ways,
eg., by considering the asymptotic expansion of the period integrals.
The geometric periods we find turn out to be most simply expressed in
terms of their derivatives, ie., in terms of the standard $K3$
periods \Kper. This is because the $K3$ periods can be written
\doubref\LSW\LY\ in terms of ordinary hypergeometric functions, and
this is very convenient for analytic continuation. Explicitly, we
find for the geometric periods $\varpi\equiv (\s,\s_{D1},\s_{D2})^t$
(up to integral changes of basis):
$$
{\del\over\del u}\,\varpi\ \equiv\
\pmatrix {\omega\cr \omega_{D1} \cr \omega_{D2}}\
=\ \pmatrix {{\zeta_0}^2 \cr \zeta_0\zeta_1-i {\zeta_0}^2 \cr
{\zeta_1}^2-2i\zeta_0\zeta_1-{\zeta_0}^2}\ ,
\eqn\kthreegeomper
$$
where
$$
\eqalign{
\zeta_0\ &=\
(u^3-1)^{1/12}\,{}_2F_1\Big(\Coeff1{12},
\Coeff7{12},1,\Coeff1{1-u^3}\Big)\cr
\zeta_1\ &=\ i\gamma\,
{}_2F_1\Big(\Coeff1{12},\Coeff1{12},\Coeff12,{1-u^3}\Big)\ ,
\ \ \ {\rm with}\ \
\gamma\equiv  {\Gamma(1/12)\Gamma(5/12)\over2\pi^{3/2}}\ .
\cr
}$$
At weak coupling, the precise forms of the leading terms are:
$$
\eqalign{
\s(u)\ &=\ 2\sqrt u+ O(u^{-5/2})\cr
\s_{D1}(u)\ &=\ {3i\over\pi}\sqrt u (\log [12u]-2)+ O(u^{-5/2})   \cr
\s_{D2}(u)\ &=\ - {9\over2\pi^2}\sqrt u (\log[12u]^2-4\log[12u]+8) +
 O(u^{-5/2})\ .
}\eqn\infper
$$
The periods turn out to obey the following identity:
$$
u(\s)\ =\ -\Coeff{72}{\pi^2}\,\varpi\cdot C\cdot\varpi\ , \ \
\ {\rm with\ intersection\ form}\ \ \ C \equiv
\pmatrix{ 0 & 0 & 1 \cr 0 & -2 & 0 \cr 1 & 0 & 0 \cr  }\ ,
\eqn\funnyrel
$$
which reflects the algebraic dependence of the $K3$ periods.
Using the well-known formulas for the analytic continuation
of hypergeometric functions, and \funnyrel\ for fixing the
integration constants, we find for the periods near the singularity
$w\equiv (u-1)\to 0$:
\def\frac#1#2{{#1\over#2}}
$$
\eqalign{
\s(w)\ &=\ \Big(\frac{{{\gamma }^2}}{4}+
        \frac{36}{{{\pi }^2} {{\gamma }^2}}\Big)+
    \frac{1}{4}  w {{\gamma }^2}-
     \frac{2  {w^{3/2}}}{{\sqrt{3}} \pi }+O(w^2)
      \cr
\s_{D1}(w)\ &=\ \Big(\frac{i  {{\gamma }^2}}{4}-
       \frac{36 i }{{{\pi }^2} {{\gamma }^2}}\Big)+
     \frac{1}{4} i  w {{\gamma }^2}+O(w^2)
     \cr
\s_{D2}(w)\ &= -\Big(\frac{{{\gamma }^2}}{4}+
        \frac{36}{{{\pi }^2} {{\gamma }^2}}\Big)-
   \frac{1}{4}  w {{\gamma }^2}-
     \frac{2  {w^{3/2}}}{{\sqrt{3}} \pi }+O(w^2)
}\eqn\oneper
$$
Note their power-like, non-logarithmic behavior near the strong
coupling singularity. From \infper\ and \oneper\ we infer the following
monodromy matrices (acting on $\varpi$) associated with $u=\infty,1$:
$$
\eqalign{
M_\infty\ &=\
\pmatrix{ -1 & 0 & 0 \cr -3 & -1 & 0 \cr -9 & -6 & -1 \cr  }\ \equiv\
T^3\cr
M_1\ &=\
\pmatrix{ 0 & 0 & -1 \cr 0 & 1 & 0 \cr -1 & 0 & 0 \cr  }\ \equiv\ S\ ,
\qquad {\rm with}\ \ {S}^2={\bfone}\ ,
}\eqn\monodr
$$
and similar (via conjugation) for the other two strong coupling
singularities. We observe an ``S-duality'' at $u=1$, associated with
$\tau_{K3}(0)\equiv {\del_w\s_{D1}\over\del_w\s}|_{w=0}=i$. Note also
that in contrast to an elliptic curve, where the point of enhanced
$\ZZ_2$ symmetry corresponds to a smooth curve, this point corresponds
here to a singular $K3$ with vanishing period $\s+\s_{D2}|_{(w=0)}=0$.

\subsec{The quantum FI coupling}

We are now equipped to compute correlation functions.
The 4-point function is particularly easy to compute, because it is
encoded in the PF system. One can actually obtain the rigid coupling
directly from the rigid PF equation, which is even
simpler. Concretely, the classical coupling is defined by
$$
C_{uuuu}\ =\ -\int_X\Omega_{4,0}\,\wedge\, {\del_u}^4\, \Omega_{4,0}\ .
$$
In the rigid limit, and integrating out two dimensions exactly as in
\KLMVW, the holomorphic 4-form turns into the meromorphic 2-form on the
$K3$, so that
$$
C_{uuuu}\ =\ \int_{K3}(\del_u\lambda_{2})\,\wedge\, {\del_u}^3\,
\lambda_{2}
\ =\ \int_{K3} \Omega_{2,0}\,\wedge\, {\del_u}^2\, \Omega_{2,0}\ .
$$
where we used \derel. We can then use that the periods \Kper\
satisfy the hypergeometric system of type
${}_3F_2(\coeff16,\coeff16,\coeff16;\coeff13,
\coeff23;u^3)$, which is of the form $L_{K3}=\sum f_k(u) {\del_u}^k$.
Following a similar strategy as in \philip, by defining
$$
V^{(k)}(u)\ \equiv\ \int_{K3} \Omega_{2,0}\,\wedge\, (\del_u)^k\,
\Omega_{2,0}
=\ \omega (\del_u)^k \omega_{D2} + \omega_{D2} (\del_u)^k \omega +
\kappa\, \omega_{D1} (\del_u)^k \omega_{D1}
$$
(where the self-intersection  number $\kappa$ is any
constant), we know that
$$
0=\sum f_k(u) V^{(k)}(u) \ \equiv\
(1-{u^3}) V^{(3)}-\Coeff{9}{2} V^{(2)} {u^2}-\Coeff{13}{4} V^{(1)}
u-\Coeff1{8}{V^{(0)}}\ .
$$
On the other hand, $V^{(0)}=V^{(1)}=0$ (due to the algebraic dependence
of the periods) and $V^{(3)}={3\over2}\del_u
V^{(2)}-{1\over2}{\del_u}^2 V^{(1)}$, whence
$3 u^2V^{(2)}+({u^3}-1) (V^{(2)})^{\prime }=0$. This has as solution
$$
V^{(2)}(u) \ \equiv\ C_{uuuu}\ =\ {1\over u^3-1}\ ,
$$
something one might have guessed beforehand. Using
$
u\sim\s^2 + O(\s^{-4})
$
the coupling in the flat variable $\s$ thus is
$$
C_{\s\s\s\s}\ =\ {1\over u(\s)^3-1}
\Big({\del u(\s)\over\del \s}\Big)^4\ \sim\
{1\over \s^2}+ O(\s^{-8})\ .
$$
We are however interested in the 3-point coupling $C_{\a\a} \equiv
C_{112}$; from factorization of amplitudes \factor\ we have that
$C_{\s\s\s\s}(\s)\sim(C_{\s\s}(\s))^2$ which gives $C_{\s\s}\sim
1/\s+\dots$. By \wdpr, this can be two times integrated to finally give
the twisted chiral potential. In fact, we find that
$$
\mathboxit{
\tilde W(\s)\ =\ \s_{D1}(\s) 
}
\eqn\thisisnice
$$
up to integration constants. This reflects the result \doubref\PM\KLRY\
for the non-rigid case, that the cubic couplings $C_{112}$ are given by
second derivatives of the ``middle'' periods with respect to flat
coordinates. Comparing to \taudef\ it thus follows that the
Fayet-Iliopoulos coupling $\tau$ in \taueff\ coincides with the modular
parameter of the auxiliary $K3$ surface, up to a possible integration
constant:
$$
\eqalign{
\tau(\s)\ &\equiv\ \tilde W'(\s)\ =\ \tauk(\s)  \cr &\!\!=
\tau_0\!-\! {6\over2\pi i}\Big[
\log({\sigma\over\mu})-\Coeff{25}{432}{\beta\over\sigma^6}-
   \Coeff{9449}{497664}\Big({\beta\over\sigma^6}\Big)^2\!\!-\!
   \Coeff{4521065}{483729408}\Big({\beta\over\sigma^6}\Big)^3
\!+\!\dots\!\Big].
}\eqn\mainresult
$$
where the bare coupling is $\tau_0={3i\over2\pi}\log3$. Above,
we inferred the normalization, ie.\ the factor $N=6$, from
$u^3=j(\tau)$, and we have reinstated the dependence on $\beta\equiv
e^{2\pi i\tau_0}\mu^6$. Because $N$ is positive, the theory is
indeed asymptotically free.

Note that $\tilde W(\s)$, being given by a period
integral,\foot{Superpotentials identified with period integrals
recently came up in $N=1$ SQCD \MQCD.} transforms non-trivially under
monodromies induced by looping around the singularities in the moduli
space. It is thus a section and not a function, as it is usual for
holomorphic quantities in supersymmetric theories. The identification
$\tilde W(\s)=\s_{D1}$ also makes sense from the point of view of
central charges: in analogy to $N=2$ gauge theory in 4d, one would be
tempted to write for the central charge of the superalgebra:
$Z=n\s+m\s_{D1}+k \s_{D2}$, and it is indeed well-known \Z\ that in 2d
the superpotential figures in the central charge.

 Formally, $n,m,k$
correspond to quantum numbers associated with $D2,D4,D6$ branes of the
type IIA string wrapped around $2,4,6$-cycles of the 4-fold,
respectively. In this sense the singularity at $u=1$ would be
associated with a massless state with $(n,m,k)=(1,0,1)$. However, such
an interpretation appears to be problematical in $d\leq3$, because of
divergences the BPS mass formula does not make much sense for fields
charged under local gauge currents \AHISS.

\subsec{Rigid Special Geometry}

One might a priori expect some generalization of rigid special geometry
\CAF\ to constrain the K\"ahler potential $K(\S,\bar\S)$.\foot
{Of course, the K\"ahler potential is not protected from quantum
corrections, so this is to be taken {\it cum} {\it grano} {\it salis}.}
Indeed there is a natural expression for a K\"ahler potential for any
$d$-fold, given by $e^{-K}=\int \Omega\wedge\bar\Omega$. In the rigid
limit, the natural expression for our rigid K\"ahler potential thus is
$$
\eqalign{
K(\s,\bar\s)\ &=\
\varpi\cdot C\cdot\bar\varpi\
=\ \s\,\bar\s_{D2}+
\bar \s\,\s_{D2}-2\s_{D1}\bar \s_{D1}
\cr&
\sim \s\bar\s+\s\bar\s (\log(\s)+\log(\bar\s))^2+\dots\ ,
}\eqn\kahler
$$
which leads to the following metric:
$$
{g(\s,\bar\s)\ \equiv\  \del\bar\del K(\s,\bar\s)\ =\
\big({\rm Im}\,\tau(\s)\big)^2\ ,
}\eqn\metric
$$
where $\tau$ is the FI coupling as above. Note that the metric \metric\
is invariant under discrete theta-shifts, $\tau\to\tau+m$, despite of
the unusual asymptotic form of the K\"ahler potential \kahler. Indeed
$K$ looks similar but different as compared to the familiar one-loop
K\"ahler potentials of the $C\IP^1$ model or the $SU(2)$ gauge theory.
The latter has the form \Jan\ $K_{1-loop} = \int_{e^2} \Coeff{dy}y
\log(1+{\s\bar\s\over y}) = {\rm Li}_2({\s\bar\s\over e^2})\sim
\log(\s)\log(\bar\s)$, which lacks the prefactor $\s\bar\s$ of our
K\"ahler potential (as it must be on dimensional grounds; in contrast,
our field $\s$ has scaling dimension equal to one). At any rate, the
important structure is the presence of quadratic logarithms,
and this is a characteristic property of ``rigid'' 2-folds.

The generalization of the well-known special geometry relation,
$R\sim 2 g^2  + e^{K}g^{-1}C_{111}\overline{C_{111}}$ \spec, for
3-folds
to $4$-folds has been discussed in \GMP. In the rigid limit, this
generalization takes the form
$
R_{\s\bar\s\s\bar\s} \sim (R^{-1})^{\s\bar\s\s\bar\s}C_{\s\s\s\s}
\overline{C_{\s\s\s\s}}
$,
which in view of the factorization property \factor\ is equivalent to
$$
R_{\s\bar\s\s\bar\s}\ =\ 2 C_{\s\s}\overline{C_{\s\s}}\ .
\eqn\specgeo
$$
Using the couplings $C_{\s\s}=\tau'$ as well as $R=g\bar\del g^{-1}\del
g$, it is trivial to check that this relation is indeed satisfied,
showing consistency of our results. Note also that by taking
derivatives of the metric \metric\ and using \factor, we get among
other terms the following term: $\s\s\del\s\bar\del\bar\s
C_{\s\s\s\s}$. This shows that the holomorphic four-point correlator
\corrs\ indeed contributes to the non-holomorphic K\"ahler potential.

\chapter{Discussion}

\subsec{4-Flux and matter}

In our treatment of the example we have so far been tacitly neglecting
the fact that the FI potential $\tilde W$ really is multiplied by an
integer number, the 4-flux $\nu$ in \wdpr. The mirror symmetry
computation we have done concerns only the correlator
$C_{112}=\langle\o^{(1)}\o^{(1)}\o^{(2)}\rangle$, and is insensitive to
the overall factor $\nu$.

The superpotential $\tilde W$ \thisisnice\ that we have been computing
thus only, but then necessarily appears if we turn on the 4-flux (if we
were free to do so, see below).  This has various implications: first,
since $TrQ$ is proportional to $\nu$, this is, from a field theory
point of view, equivalent to switching on charged matter. That is, in
the effective lagrangian we cannot distinguish the logarithmic term
${6\nu\over 2\pi i}\Sigma\log\Sigma$ in $\tilde W$ that we get from
geometry, from a field theory one-loop term ${TrQ\over 2\pi
i}\Sigma\log\Sigma$, generated by integrating out massive chiral matter
multiplets. In other words, the effective theory behaves exactly in the
way as there would be charged matter, although geometrically we did not
put any in.

This may also have a bearing on chiral matter in
$N=1$ theories in four dimensions, where there generically is a
Green-Schwarz anomaly cancelling term of the form: ${\cal L}_{GS}=c\,
(B\wedge F)$ with $c\equiv {\rm Tr} Q$. In F-theory compactified on a
4-fold, this term is obtained from the following formal coupling
\FMS\ in twelve dimensions: ${\cal L}=A_4\wedge F_4\wedge F_4$,
where $A_4$ is a 4-form gauge field. That is, upon expanding $A_4 =
B\wedge \o^{(1)}$ and $F_4$ as in \formexpand, one obtains this
coupling with $c = \nu\, \langle\o^{(1)}\o^{(1)}\o^{(2)}\rangle_{{\rm
class}}$, where "class" denotes the classical intersection (what
figured in $d=2$ was the world-sheet corrected quantum version of
this). This means that whenever there is an anomalous $U(1)$ in
$d=4$, in $F$-theory language some 4-flux $\nu$
must to be non-zero. Turning this around, switching on $\nu$ may in
some sense {\it implement} chiral matter.

In addition it must be that $SU(2)$ is broken when we switch on $\nu$,
simply because $TrQ$ vanishes identically for any non-abelian group.
Indeed, the resulting potential $\int\!\nu\tilde W(\Sigma)\sim
\nu(\Sigma+\Sigma\log\Sigma+\dots)$ is not invariant under discrete
Weyl transformations ($\Sigma\to-\Sigma$). However only mildly so: it
just changes sign, up to an additional theta-shift. Weyl invariance of
the effective theory can therefore be restored if we simultaneously
flip the signs of $\s$ and of the symmetry breaking flux $\nu$. In this
way the theory exhibits the presence of the $SU(2)$ that was originally
built in.

\subsec{Supersymmetric vacua ?}

Another important implication of turning on 4-flux is a non-vanishing
scalar potential \Vpot, $V\sim\nu^2 |\tau(\s)|^2$, where the FI
coupling is given in terms of the $K3$ periods \kthreegeomper\ as
$\tau=\tau_{K3}\equiv\omega_{D1}(\s)/\omega(\s)$. One may wonder
whether there are any supersymmetric vacua given by $\tau=0$.\foot
{This ties together with the following condition for a supersymmetric
4-form background \beckerSquared: $ (F_4)_{a\bar bc\bar d}J^{\bar cd} =
0$, where $J$ is the K\"ahler form.  This Uhlenbeck-Yau type of
equation implies \GSW $\int J\wedge J\wedge F_4=0$, which  leads to
\PM\ $ \sum t_i t_j\nu_k \langle\o^{(1)}\o^{(1)}\o^{(2)}\rangle=0$,
where $t_i$ are special coordinates. We recognize here essentially the
condition for a vanishing FI potential.} Semiclassically, where
$\tau\sim\tau_0-{6\over2\pi i} \log\s$, supersymmetric vacua obviously
do exist.

But non-perturbatively, $\tau$ is non-zero everywhere over the moduli
space, because it is a modular parameter living in three copies of the
usual fundamental region. This means that supersymmetry must be
spontaneously broken~! One might say that this happens because the
relevant 2-cycle has non-zero ``quantum volume'' \QV\ throughout the
moduli space. One could have speculated that there would be a zero
quantum volume at the conifold point $u=1$, similar to what happens for
certain vanishing del Pezzo 4-cycles \delP, but it did not turn out
that way. Specifically we have at the conifold point $\tau=i$, which
corresponds to a non-vanishing FI parameter, $\xi=1$.
Note that in view of
\oneper, the physics at the singularity is governed by a power-like,
and not logarithmic, potential.


The story is however not that clearcut, as $\nu$ cannot, in general,
be adjusted at will. Indeed from \LCS\ one has in addition a tadpole
$B\nu^2\langle\o^{(2)}\o^{(2)}\rangle$, besides the 1-loop tadpole
$B\int I_8(R)$ \VW, and all these tadpoles need to be cancelled
together with contributions from extra 1-branes \SVW; this sometimes
even forces $\nu$ to be non-zero \refs{\witflux{,}\DM}. These
considerations involve global properties of the compactification
4-fold $X$, while we were discussing in our example only local
properties. We neglected in particular possible effects of extra
tadpole cancelling branes. It may thus well be that in our local
analysis, we did not capture an important ``global'' ingredient, like
a contribution to the vacuum energy. What we have in mind is that
there might be an extra bare FI coupling in the potential,
$$
\tilde W(\Sigma)\ \longrightarrow \tilde W(\Sigma)+\delta\,\Sigma\
\eqn\fudgeterm
$$
which shifts the vacuum energy and restores a supersymmetric vacuum. In
fact, in our computation of $\tilde W$ in \thisisnice\ there was room
for an integration constant of exactly that type, so all we can really
say from our computation is that $\tau=\tau_{K3}+\d$, though a non-zero
$\d$ would not be natural from the geometrical point of view.  Note
that adding such a constant also does not violate the special geometry
relation \specgeo.

It would appear particularly appealing to add such a term with
$\delta=-i$.\foot{For simplicity, we will set $\nu=1$ in the
following}. Namely without it, the potential $V\sim|\tau|^2$ would have
a flat direction along the lower boundary of the fundamental region
(the arc between Im$\tau=-\shalf$ and Im$\tau=+\shalf$), which would
not seem to make much sense. After adding this term, the vacuum
degeneracy is resolved, a mass gap created and there is a supersymmetry
preserving ground state located at the self-dual conifold point $u=1$
($\tau_{K3}=i$). This would be in line with the arguments of
\stro\ that say when switching on $p$-form fluxes, potentials are
generated whose minima lie on conifold points. 

The other two images of the conifold point (located at $u=(-1)^{\pm
2/3}$) correspond to $\tau_{K3}=i\pm1$, ie., to non-vanishing
theta-angles $\pm2\pi$. Physically, a non-zero theta-angle describes a
constant electric field that contributes to the vacuum energy \colem.
The point is that $\theta$ naturally lives in the domain
$-\pi\leq\theta\leq\pi$, because for $|\theta|>\pi$ the vacuum energy
can be reduced by pair creation to $|\theta|\leq\pi$. Semi-classically,
one thus defines \witphases\ an effective theta-angle in terms of a
piecewise smooth function $\tilde\theta=\theta+2\pi n$, such that
$|\tilde\theta|\leq\pi$.\foot{A natural way to restore the
$2\pi$-periodicity of the effective theta-angle in the non-perturbative
theory, would be simply go to a triple cover of the moduli
space, parametrized by $\tilde u \equiv u^3 = j(\tau_{K3})$, yielding a
smooth and $2\pi$-periodic function for $\theta(\tilde u)$. This
parametrization would be natural from the viewpoint of $K3$
geometry, but not natural from the viewpoint of a fibered $A_1$
singularity. }

In this sense there are then six semi-classical vacua given by
$\s_n=e^{2\pi i(\tau_0+n)/6}$, $n=0,..,5$, because all $\s_n$ lead to
$\tilde \theta=0$. This gives a non-zero Witten index, which precludes
a spontaneous breakdown of supersymmetry. If we add $\delta=-i$, then
we consistently have, in the same sense, also at the non-perturbative
level six vacua, coming from the three singularities at $u^3=1$, each
singularity counting two vacua because of the enhanced $\ZZ_2$ symmetry
there.

\subsec{Generalizations}

One can easily see that all fibrations of $\IP^1$ over
$\IP^2$ give the same rigid limit.
In our example, one may as well take for instance
$\IP^1\times\IP^2$ with $l_f=(1,1,0,0,0,-2)$ and
$l_b=(0,0,1,1,1,-3)$, instead of \mori.  This is completely
analogous to four dimensions, where all the fibrations
of $\IP^1$ over $\IP^1$ (the Hirzebruch surfaces $F_n$), give in the
rigid limit the same result, ie., the $SU(2)$ Seiberg-Witten curve, for
all $n$ \doubref\KKV\KMV. Thus there is an analogous universality for
the 2d rigid theories based on $\IP^2$.

Note however that since the base is two-dimensional, there is more
than only one choice for it: $\IP^2$ just corresponds to the simplest
possibility with the lowest number of parameters. The ubiquitous factor
of 3 we encounter (giving the order of splitting of the classical
singularity, the order of the PF operator, the value of $N=6\equiv
2\times 3$ and the $\ZZ_6$ symmetry of the instanton expansion) traces
back to the intersection of $c_1(\IP^2)$ with the $h_{1,1}$ class,
which is 3. For other choices of the base there will be other such
characteristic numbers, and therefore these theories will be different
from the model we have been discussing.

One can for instance also consider fibrations of $\IP^1$ over
Hirzebruch surfaces $F_n$, where the global symmetry is
$\ZZ_8=(\ZZ_2)^3$ instead of $\ZZ_6$. A new feature will be the
appearance of a second ``quantum scale'' $\tilde\beta$. Moreover, the
rigid limit will be independent only in the way $\IP^1$ is fibered over
$F_n$, but it will not be independent of $n$. One can easily see that
the resulting rigid surfaces have the form
$$
W_{K3}\ =\ z+w + {\beta\over z}+ {z^n\tilde\beta\over w} + (x^2+u)
\ =\ 0\ ,
$$
which give quartic $K3$'s for $n=0,..,3$. Obviously one has
universality only in the limit $\tilde\beta\to0$, where one recovers
the well-known SW gauge coupling, and indeed one may view these
theories as coming from fibrations of the SW geometry over a further
$\IP^1$ (whose size is governed by $\tilde\beta$). Accordingly each
of the two monopole singularities is split into two further
singularities, the scales of the two independent splittings being
$\beta$ and $\tilde\beta$.

For higher rank groups (ie., general ADE singularities fibered over
$\IP^2$), we will generically not find $K3$ surfaces. This is similar
to $d=4$, where the SW curves for $SU(n)$ are not Calabi-Yau (ie.,
$c_1\not=0$) for $n>2$. For example, fibering an $A_2$ singularity we
will end up with $W= z+w+{\beta^6\over zw}+(x^3+u x+v)=0$,
and the homogenous form of this surface in $W\IP(1,1,1,1;6)$ does not
give a $K3$. We expect it to have not one, but two holomorphic $(2,0)$
forms (and analogously $n-1$ holomorphic $(2,0)$ forms for $A_{n-1}$
singularities, so that $\del_{u_{k+1}}\lambda_2=\Omega_{2,0}^{(k)}$,
$k=1,..,n-1$).

For arbitrary groups $G$, there will be vectors of 4-fluxes and
potentials $\tilde W_i\sim (\vec\alpha_i\cdot\vec\Sigma)
\log(\vec\alpha_i\cdot\vec\Sigma)$, of which rank($G$) are independent.
Integrating the topological term ${\theta\over2\pi} (\vec F\cdot \vec
\nu)$ over 2d space time, we see that the 4-fluxes $\vec\nu$ must
indeed be quantized and dual to the instanton numbers $\vec
n={1\over2\pi}\int d^2x\vec F$. They thus naturally lie on the
corresponding weight lattice, $\vec\nu\in\Lambda_w^G$, and we expect
this also to be reflected by the intersection properties of the middle
homology. Weyl invariance (up to theta-shifts) can be restored by
simultaneously Weyl transforming $\vec\s$ and the symmetry breaking
fluxes $\vec\nu$, and in this way the effective theory exhibits the
underlying group $G$.

\chapter{Conclusions}

We have been investigating a field theory limit of type IIA string
compactification on certain 4-folds that have as compact piece a
fibration of $\IP^1$ over $\IP^2$. The effective action is given by a
$U(1)\times U(1)$ gauge theory, where one $U(1)$ factor decouples. It
is characterized by a complex Fayet-Iliopoulos coupling $\tau(\s)$
given in \mainresult, which displays a logarithmic one-loop piece plus
further non-perturbative corrections. The twisted chiral superpotential
$\tilde W(\Sigma)$ is given by the middle period of an auxiliary $K3$
surface. This surface encodes non-perturbative information about the
Coulomb branch of the effective $U(1)$ theory, and thus is the
two-dimensional analog of the SW curve in four dimensions.

Although {\it a priori} the geometrical setup of this kind of theories
appears to be similar to that of $N=2$ gauge theories in $d=4$ \SW, the
theories turn out to be remarkably different.  What we find in $d=2$
are not just the usual $SU(2)$ (or more generally ADE) gauge theories
or Grassmannian models, but theories whose structure is richer.

An important difference as compared to 4d is that the perturbative
piece of such a 2d field theory does not fully determine the
non-perturbative theory.  That is, the logarithmic one-loop term arises
from wrappings of world-sheet instantons around the fiber, and is
universal for a given fibered ADE singularity. On the other hand,
wrappings around the various classes of the base manifold give
non-perturbative corrections, and different choices for the base
provide different ``non-perturbative completions'' of the perturbative
theory. This is extra information which goes beyond the perturbative
definition of the theory, but which is required for the global
consistency of the full theory. It may simply be that one cannot well
define the 2d theory unless it is embedded in a larger consistent
framework, and, unlike as in $d=4$ where there is only one choice for
the base, this embedding turns out to be ambiguous.

Besides continuous moduli, the effective theory also possess
discrete moduli to play with, the 4-fluxes $\nu$.

If there is no 4-flux turned on, the 2d theory is essentially a
reduction of an ADE gauge theory in six dimensions on some 2-fold base.
For such a theory the holomorphic potential $\tilde W(\Sigma)$ that we
have been computing does not appear in the effective lagrangian. Still,
even for vanishing $\nu$,  there remains some structure in the theory,
namely a non-trivial K\"ahler potential and a non-trivial complex
structure moduli space.  The K\"ahler potential is not protected from
corrections, and it is thus not clear what conclusions can be drawn
from the geometric expression for it given in \kahler.

In our example,
the complex structure quantum moduli space has the form of a
non-abelian orbifold, ${\cal M}=\IH/\Gamma_1$, where $\Gamma_1$ is the
monodromy group of our problem. Strictly speaking, since there is no
good notion of a VEV in two dimensions, the moduli space should rather
be viewed as a target space of a sigma-model, to be functionally
integrated over. Remember however that a fivebrane instanton induced
potential cannot be excluded if $\nu=0$ (see section 3), and is in fact
expected to appear \doubref\superpot\KVsuppot. Such a potential would
substantially change the vacuum structure. It is not clear to us,
though, how to compute it using mirror symmetry.

More interestingly, the theory looks very different if we turn on
$\nu$: then a non-trivial potential $\tilde W(\Sigma)$ is generated,
which breaks the ADE symmetry and which reflects extra matter fields
that were not present before. At the perturbative level the theory is
simply a $U(1)$ gauge theory with some charged matter, and hence
similar to a $CP^n$ model. Non-perturbatively, a non-vanishing 4-flux
leads to a spontaneous breakdown of supersymmetry, unless a bare
coupling is added by hand. In this case a mass gap appears and
supersymmetric ground states are naturally associated with
singularities in the moduli space. At large distances the K\"ahler
potential becomes then irrelevant, and the theory becomes a topological
field theory with Chern-Simons lagrangian ${\cal L}=\nu\,\tilde
W(\Sigma)$. It can probably be interpreted along the lines of \grass\
as some kind of abelian WZW model at ``level $\nu$''.

We see that switching on the 4-flux is a quite drastic operation, and
resembles a bit to switching on a shift vector in orbifold
compactifications. It would seem worthwhile to investigate the r\^ole
of the 4-flux also from that perspective.

\ack

I would like to thank
M.\ Bershadsky,
J.\ de Boer,
M.\ Douglas,
S.\ Ferrara,
E.\ Kiritsis,
A.\ Hanany,
P.\ Mayr,
R.\ Minasian,
Y.\ Oz,
S.\ Seiberg,
S.\ Stieberger,
S.\ Theisen and especially C.\ Vafa
for sharing their insights.

\refout
\end